\documentclass[prl,twocolumn,showpacs,superscriptaddress]{revtex4}
\usepackage{graphicx}
\usepackage{amsmath}
\usepackage{amssymb}
\usepackage{comment}
\usepackage{psfrag}

\begin{document}

\title{All-optical non-demolition measurement of single-hole spin
in a quantum-dot molecule}

\author{F. Troiani}
\email[]{filippo.troiani@uam.es} \affiliation{Departamento de
F\'\i sica Te\'orica de la Materia Condensada, Universidad Aut\'onoma de
Madrid, 28049 Madrid, Spain}
\author{I. Wilson-Rae}
\affiliation{Departamento de F\'\i sica Te\'orica de la Materia
Condensada, Universidad Aut\'onoma de Madrid, 28049 Madrid, Spain}
\author{C. Tejedor}
\affiliation{Departamento de F\'\i sica Te\'orica de la Materia
Condensada, Universidad Aut\'onoma de Madrid, 28049 Madrid, Spain}

\date{\today}

\begin{abstract}

We propose an all-optical scheme to perform a non-demolition
measurement of a single hole spin localized in a quantum-dot
molecule. The latter is embedded in a microcavity and driven by two
lasers. This allows to induce Raman transitions which entangle the
spin state with the polarization of the emitted photons. We find
that the measurement can be completed with high fidelity on a
timescale $T \sim 10^{2}$~ps, shorter than the typical $T_2$.
Furthermore, we show that the scheme can be used to induce and
observe spin oscillations without the need of time-dependent
magnetic fields.

\end{abstract}


\maketitle

The capability of encoding and manipulating information at the
single-spin level represents a key challenge for semiconductor-based
spintronics and quantum-information~\cite{zutic,nielsen}. A reliable
read-out of an individual-spin state is likely to require the
measurement to be repeatable. This calls for it to be
non-destructive and carried out on timescales shorter than those
characterizing spin decoherence~\cite{meunier,liu}. In this respect,
optical manipulation of single carriers in quantum dots (QDs) is
specially attractive due to the orders of magnitude separation
between optical timescales and those associated to the intrinsic
spin dynamics~\cite{troiani:03}. While most of the attention has
been centered in the past on electron spin, it can be argued that
hole spin could offer novel alternatives. Along these lines, it has
been noted that the decoherence due to hyperfine interactions is
suppressed compared to that affecting the electron~\cite{hrelax}.

\begin{figure}
\begin{center}
\includegraphics[width=\columnwidth]{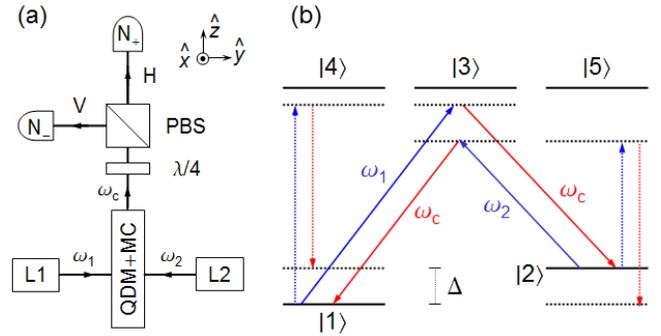}
\caption{(Color online). (a) Schematic diagram of a possible
experimental setup: photocounts $N_+$ ($N_-$) are recorded at the
right (left) detector and the measurement outcome ($|\!\!\uparrow
\rangle$ or $|\!\!\downarrow\rangle$) is decided based on whether
($N_+\gtrless N_-$).  (b) Level scheme corresponding to each of the
two subspaces, ``$+$'' and ``$-$''. Optical transitions are induced
by two lasers (blue arrows) with frequencies $\omega_1$ and
$\omega_2$ linearly polarized along $\hat{x}$, and two degenerate
cavity modes (red) with frequency $\omega_c$.  The scheme relies on
Raman transitions between the states $ |1\pm\rangle $ and $
|2\pm\rangle $, that involve $\sigma_{\pm}$ radiation (solid
arrows). All deleterious virtual processes that involve the emission
of $\sigma_{\mp}$ photons (an example of which is given by the
dotted arrows) are very off-resonant.\vspace{-1cm}} \label{fig1}
\end{center}
\end{figure}

Here we propose a novel technique to perform a fast and robust
non-demolition measurement of single hole spin in a QD-microcavity
(MC) system. To illustrate its merit, we also discuss how it could
be used to study the spin decoherence with a photon-correlation
experiment. The basic idea is to exploit virtual Raman transitions
that entangle the spin ($|\!\!\uparrow \rangle$ or $|\!\!\downarrow
\rangle$) with the polarization ($\sigma_+$ or $\sigma_-$) of the
photons emitted into the cavity.  The semiconductor heterostructure
we consider consists of two self-assembled QDs, coherently coupled
with each other and embedded in a high-Q optical MC. The quantum-dot
molecule (QDM) is doped with an excess hole~\cite{stinaff}, and its
lowest-energy trion transition is strongly coupled to a pair of
degenerate cavity modes with frequency $\omega_c$, damping constant
$\kappa$, and polarizations $\sigma_{\pm}$ \cite{microp}. In the
absence of a magnetic field, the ground state of the hole is doubly
degenerate and each of the two eigenstates of its spin along the
optical axis [$\hat z$ in Fig.~\ref{fig1}(a)] couples to a different
set of trion states.  The system's dynamics is driven by two
linearly polarized lasers ($1$ and $2$) with frequencies $\omega_1$
and $\omega_2$. The photons emitted by the cavity are sorted out
with a $\lambda/4$ phase shifter followed by a polarizing beam
splitter. The photons with polarization $\sigma_+$ ($\sigma_-$) are
finally sent to the right (left) detector where photocounts $N_+$
($N_-$) are recorded. The outcome of the spin ($J_z$) measurement is
decided based on whether $N_+\gtrless N_-$, rather than on the
presence versus absence of photons in a given mode, and only events
in which the measurement outcome satisfies $N_+ \neq N_-$ are
post-selected. This strategy makes our scheme intrinsically
resilient against photon loss and detector inefficiency. In
addition, the asymmetry of the QDM allows to use laser frequencies
that are out of resonance with the cavity, and thus to spectrally
resolve the output from the light scattered by the heterostructure.

A typical QDM is formed by two vertically-stacked QDs. There, the
combined effect of strain and effective-mass asymmetry strongly
suppresses the hole-state hybridization, while allowing the
formation of molecular-like bonding and antibonding states for
electrons~\cite{bester:04}.  Thus, we assume that the heavy hole
($J_z=\pm 3/2$) remains localized either in the larger dot ($L$) or
in the smaller one ($S$), while the electron ($S_z=\pm 1/2$) bonding
state ($B$) is significantly delocalized over the two.  At low
temperatures ($T<5$K) and for near resonant interaction with the
laser and cavity fields, we can restrict ourselves to a ground-state
manifold: $ |1\pm \rangle\equiv d^\dag_{\uparrow/\downarrow L}
|0\rangle , |2\pm \rangle\equiv d^\dag_{\uparrow/\downarrow S}
|0\rangle $, and an excited-state (trion) manifold: $|3\pm
\rangle\equiv c^\dag_{\downarrow/\uparrow B}
d^\dag_{\uparrow/\downarrow L} d^\dag_{\uparrow/\downarrow S}
|0\rangle$, $|4\pm \rangle\equiv c^\dag_{\uparrow/\downarrow B}
d^\dag_{\uparrow/\downarrow L} d^\dag_{\downarrow/\uparrow S}
|0\rangle$, $|5\pm \rangle\equiv c^\dag_{\uparrow/\downarrow B}
d^\dag_{\downarrow/\uparrow L} d^\dag_{\uparrow/\downarrow S}
|0\rangle$ [see Fig.~\ref{fig1}(b)]. The latter comprises the trion
states formed by one heavy-hole per dot and one electron in the
bonding state that are optically active, and is separated from
$|1\pm \rangle$ by an optical energy difference $\hbar \omega_T$.
Due to the QDM asymmetry the ground state manifold presents an
energy splitting $\hbar\Delta$, which sets the largest energy scale
relevant for our purposes.

We note that {\it the interaction with the optical field does not
mix the ``$+$'' and ``$-$'' states of the QDM}. Thus, aside from
incoherent spin-flip processes, the dynamics of the system during
the spin measurement will satisfy the QND back-action evasion
criterion. On the other hand, Raman transitions between the $
|1\pm\rangle $ and $|2\pm\rangle $ states [solid arrows in
Fig.~\ref{fig1}(b)] give rise to a precise correlation between the
polarization of the emitted cavity photon ($\sigma_{\pm}$) and the
spin orientation ($ \uparrow \!\!/\!\! \downarrow $). The
cornerstone of the scheme is to choose the laser frequencies so that
the two photon resonance condition for these transitions is met --
i.e. $\omega_1 \approx \omega_c + \Delta$ and $\omega_2 \approx
\omega_c - \Delta$ -- while undesired processes, leading to the
emission of anticorrelated photons, are kept very off-resonant. A
Raman transition mediated by laser $1$ ($2$) transfers the hole from
dot $L$ ($S$) to dot $S$ ($L$) and creates a cavity photon. Thus,
the combined action of both lasers is equivalent to a cycling
transition between $|1\pm\rangle$ and $|2\pm\rangle$ that allows to
amplify the single spin to be measured into a many photon state.

In order to study its interaction with the radiation field, we treat
the laser driven QDM coupled to the MC as an open quantum
system.  We apply a time-dependent canonical transformation
$\mathcal O \rightarrow e^{s(t)} \mathcal O e^{-s(t)}$ defined by
\begin{equation*}
s(t)=-i \!\!\sum_{\zeta=\pm;\,n=1,2} \left[ \tilde\omega_n t -
\frac{(-1)^n}{2} \phi(t) \right] \sigma_{nn}^{(\zeta)} -
\frac{\tilde\omega_n t}{2} a_\zeta^\dag a_\zeta^{\vphantom{\dag}} ,
\end{equation*}
where: $\sigma_{mn}^{(\zeta)}\equiv | m \zeta \rangle \langle n \zeta
|$, $a_\pm$ is the annihilation operator for the $\sigma_{\pm}$
polarized cavity mode, $\tilde\omega_{1/2} \equiv
(3\omega_{1/2}+\omega_{2/1})/4$, and
$\phi(t)\!\equiv\!\sum_{n,m}^2\int_0^t\!\mathrm{d}\tau
\Omega_{n}^{(m)}(\tau)/[(-1)^n\delta_T +
  (3-4\delta_{n,m})\tilde\Delta]$. Here we also introduce
$\tilde\Delta\equiv(\omega_1-\omega_2)/2$,
$\delta_T\equiv\omega_1+\omega_2+\Delta-2 \omega_T$, and the Rabi
frequencies $\Omega_{1/2}^{(1)}(t)$ [$\Omega_{1/2}^{(2)}(t)$] for the
transitions between $|1/2\pm \rangle$ and the trion states induced by
laser $1$ ($2$).  In this representation the system's Hamiltonian is
given by  $H(t)=H_0(t)+V(t)+H_T$, with ($\hbar\equiv1$):
\begin{equation*}
H_0(t)= \sum_{\zeta=\pm}
\left[\delta_s - \dot\phi(t)/2\right]
\left[\sigma_{22}^{(\zeta)} - \sigma_{11}^{(\zeta)}\right] - \delta_c
a_\zeta^\dag a_\zeta^{\vphantom{\dag}} ,
\end{equation*}
\begin{align}
\label{eq:H}
V(t)=&  \sum_{\zeta=\pm} \sum_{n=1,2} g_n e^{i \frac{(-1)^{n-1}}{2}
[\tilde\Delta t + \phi(t)]} \left\{ \sigma_{3,n}^{(\zeta)}
\left[ \alpha(t) + a_\zeta \right]  \right. \nonumber\\
& + \left. \sigma_{n+3,n}^{(\zeta)} \left[ \alpha(t) +
a_{-\zeta} \right] \right\} + \mathrm{H.c.},
\end{align}
and $H_T = -\delta_T/2 \sum_{\zeta=\pm}\sum_{n=3}^5
\sigma_{nn}^{(\zeta)}$; where we introduce $\alpha(t)\equiv [
\Omega_1^{(1)}(t) e^{-i \tilde\Delta t} + \Omega_1^{(2)}(t) e^{i
\tilde\Delta t}]/2 g_1$, the detunings $\delta_s\equiv
(\Delta-\tilde\Delta)/2$, $\delta_c\equiv
(\omega_1+\omega_2)/2-\omega_c$, and the QDM-MC couplings
$ g_{1/2} \! > \! 0 $.  In addition there are dissipative contributions
associated to the cavity losses and to the spontaneous emission of the
trion into leaky modes. Their respective Liouvillians, $\mathcal{L}_C$
and $\mathcal{L}_T (t)$, are of the Lindblad form with collapse
operators given by: $\sqrt{\kappa} a_\pm$ and $ c_{\pm}(t) =
\sqrt{\Gamma}\sum_{n}^2 (g_2/g_1)^{n-1} e^{i(-1)^{n}[\tilde\Delta t +
\phi(t)]/2} [\sigma_{n,3}^{(\pm)} + \sigma_{n,n+3}^{(\mp)}]$. Here
$ \Gamma $ is the spontaneous-emission rate to the state $ | 1 \pm \rangle$.

In the virtual-Raman regime of interest, $|\delta_T| \lesssim
|\tilde\Delta|$, with the latter much larger than all the other
frequency scales in $H(t),\mathcal{L}_{T/C}$. This warrants an
adiabatic elimination of the trion-ground state
coherences~\cite{cohen,gardiner}. To this effect, we decompose the
density matrix of the QDM-MC system ($\rho$) into a relevant part $
\mathbb{P}\rho = P_0 \rho P_0 + P_T \rho P_T $ and an irrelevant one
$[\mathbb{I}-\mathbb{P}]\rho (t)$ -- where $P_0$ ($P_T$) is the
projector onto the ground state (trion) manifold.  We subsequently
eliminate the irrelevant part, introduce a formal parameter
$\lambda$ such that $\delta_T \to \lambda \delta_T$ and
$\tilde\Delta \to \lambda \tilde\Delta$, and consider the asymptotic
expansion of $\mathbb{P}\dot\rho$ as $\lambda \to
\infty$~\cite{gardiner}. To the lowest non-trivial order
($1/\lambda$), $ \mu \equiv P_0 \rho P_0 $ obeys a closed evolution
generated by $\mathcal{L}_{C}$ and the effective Hamiltonian
\begin{align}
H_\mathrm{eff}=&\sum_{\zeta=\pm}\delta_s\sigma_z^{(\zeta)} -
\left\{\delta_c-\Delta_c-\Delta_s\left[\sigma_z^{(\zeta)} +
  \sigma_z^{(-\zeta)}\right]\right\}a_\zeta^\dag
a_\zeta^{\vphantom{\dag}}\nonumber\\ & + \!\left\{\left[\frac{g_2
    |\Omega_1^{(1)}(t)|}{\delta_T + \tilde\Delta}\sigma_+^{(\zeta)} +
  \frac{g_1 |\Omega_2^{(2)}(t)|}{\delta_T - \tilde\Delta}
  \sigma_-^{(\zeta)}\right]a_\zeta^\dag +
\mathrm{H.c.}\!\right\} ,
\nonumber
\end{align}
where we have introduced Pauli matrix notation for the orbital
pseudospin $\sigma_+^{(\pm)}\equiv\sigma_{21}^{(\pm)}$,
$\Delta_{c/s}\equiv g_2^2/(\delta_T+\tilde\Delta)\pm
g_1^2/(\delta_T-\tilde\Delta)$, and we have chosen
$\Omega_{1/2}^{(1/2)}(t)/|\Omega_{1/2}^{(1/2)}(t)|=e^{\mp
i\phi(t)}$. On the other hand we find that the spontaneous emission
only contributes to order $1/\lambda^2$. In terms of the physical
parameters this corresponds to a correction to the evolution
generated by $H_\mathrm{eff}$ of relative order $\Gamma/\Delta\ll
1$. The above treatment is valid provided $|\Omega_1^{(1/2)}(t)|^2
\ll (n |\tilde\Delta|-|\delta_T|)^2$ and $\kappa, \Gamma,
|\delta_{c/s}|, |\dot\Omega_1^{(1/2)}(t)/\Omega_1^{(1/2)}(t)| \ll |n
|\tilde\Delta|-|\delta_T||$ are satisfied for $n=1,3$; where we
assume $g_2 < g_1 < |\Omega_1^{(1/2)}(t)|/2$ and that the typical
cavity occupancies are at most of order unity.  We take laser and
cavity frequencies so that $ \delta_{s} \! = \! 0 $, $ \delta_{c} \!
= \! \Delta_c $, $ \Delta_s \! = \! 0 $. The asymmetry of the
molecule ensures $|n |\tilde\Delta| - |\delta_T|| \sim \Delta $ ($n
\!\!=\!\!1,3$).
 The relative
intensities of the two lasers are chosen so that the coefficients of
$\sigma^{(\pm)}_{+}$ and $\sigma^{(\pm)}_{-}$ are equal. This yields
$\tilde{H}_\mathrm{eff}=-\tilde\Omega(t)/2 \sum_\zeta
\sigma^{(\zeta)}_x (a_\zeta^\dag + a_\zeta^{\vphantom{\dag}})$ with
$\tilde\Omega(t)=|\Omega_{1}^{(1)}(t)|(g_1^2-g_2^2)/g_2 \Delta$. The
lasers are switched on at $t=0$ so that for negative times
$\tilde\Omega(t)=0$ with the cavity modes in the vacuum.

As will be borne out below, to analyze the measurement process it is
useful to consider the time evolution conditioned upon having no
photocounts detected~\cite{gardiner}: $ \dot\mu \!= \!\sum_{\zeta=\pm} -i [
\tilde{H}_\mathrm{eff} ,\mu ] + \mathcal{L}_C^{(\zeta)}(\eta)\mu \,$
with $\mathcal{L}^{(\zeta)}_C(\eta)\mu
=(\kappa/2)[2(1-\eta)a_\zeta^{\vphantom{\dag}} \mu a_\zeta^\dag -\{
a_\zeta^\dag a_\zeta^{\vphantom{\dag}}, \mu \}]$. Here $\eta$ is the
collection efficiency times the efficiency of the detectors and for
$\eta\!=\!0$ one recovers the standard time evolution. This equation
for $\mu(t)$ has the following solution:
\begin{align}\label{eq:mu}
\mu(t)= & \sum_{\zeta=\pm} \sum_{\nu=0}^3 e^{-\xi_\nu \kappa P(t)} e^{-i p(t) \sigma^{(\zeta)}_x
\left(a_\zeta^\dag + a_\zeta^{\vphantom{\dag}}\right)} |0 \rangle
\langle 0 |_{+} \nonumber \\
&\otimes |0 \rangle \langle 0 |_{-}\otimes \frac{
\mathrm{Tr}[\sigma^{(\zeta)}_\nu \mu(0)]}{2}\sigma^{(\zeta)}_\nu
e^{i p(t) \sigma^{(\zeta)}_x \left(a_\zeta^\dag +
a_\zeta^{\vphantom{\dag}}\right)}\,
\end{align}
where $\xi_{0/1}\equiv \eta$, $\xi_{2/3}\equiv 2-\eta$, and
$P(t)=\int_0^t \!\mathrm{d}\tau p(\tau)^2$ with $p(t)$ satisfying:
$2\dot p = -\tilde\Omega(t) - \kappa p$, $p(0)=0$. Here $ |0 \rangle
\langle 0 |_{\pm}$ are the vacuum states for the cavity modes,
$\sigma^{(\zeta)}_0 \equiv \sigma^{(\zeta)}_{11}+
\sigma^{(\zeta)}_{22}$, and $\sigma^{(\zeta)}_\nu$ with $\nu \! > \!
0$ correspond to the Pauli matrices. The profiles of the laser
pulses are chosen so that $\tilde\Omega(t)=\tilde\Omega_0 [
e^{-8(t/t_s-1)^2}\Theta(t_s-t) + \Theta(t-t_s)]$, where $t_s$ is a
switch-on time. We choose $1/\Delta \ll t_s \ll 1/\tilde\Omega_0$ so
that $t_s\tilde\Omega_0$ is a small parameter and one can keep only
the zeroth order. This corresponds to $\kappa P(t)= p_0^2 f[\kappa
t/2]$ where $f[x]\equiv 2x + 1 - (2-e^{-x})^2$ and $p_0 \equiv
\tilde\Omega_0/\kappa$, which specifies $\mu(t)$.

If we now
consider in Eq.~(\ref{eq:mu}) $\eta=0, \, t \to \infty$ we obtain the
steady state to which the system converges starting
from a given initial condition for the hole $\rho_h$:
\begin{align}\label{eq:sstate}
&\mu_\mathrm{ss}=\!\!\sum_{\xi=\pm} \!|\!\!\uparrow\!\xi\rangle
\langle \uparrow\!\xi | \rho_h |\!\!\uparrow\!\xi\rangle \langle
\uparrow\!\xi | \!\otimes\! |\!-\!\xi i p_0 \rangle \langle
-\xi i p_0 |_{\scriptscriptstyle +} \!\otimes\! |0\rangle
\langle 0|_{\scriptscriptstyle -} \nonumber \\
&+|\!\!\downarrow\!\xi\rangle \langle \downarrow\!\xi | \rho_h
|\!\!\downarrow\!\xi\rangle \langle \downarrow\!\xi | \!\otimes\!
|0\rangle \langle 0|_{\scriptscriptstyle +} \!\otimes\!
|\!-\!\xi i p_0 \rangle \langle -\xi i p_0
|_{\scriptscriptstyle -}
\end{align}
Here we have defined
$|\!\!\uparrow\!\!/\!\!\downarrow\!\xi\rangle\equiv(|2\pm\rangle +
\xi |1\pm\rangle)/\sqrt{2}$ and introduced the cavity coherent
states $|\lambda\rangle_\pm$. We note the perfect correlation
between the initial state of the spin and the polarization of the
cavity mode that does not remain in the vacuum, and its independence
from the initial orbital state of the carrier. In addition we find
that the eigenstates of the orbital pseudospin $\sigma_x$ become
correlated with the phase quadratures of the cavity fields. Thus if
homodyne detection is performed for both circular polarizations one
can also measure the orbital state of the hole in the $|2\pm\rangle
\pm |1\pm\rangle$ basis.

\begin{figure}
\begin{center}
\includegraphics[width=\columnwidth]{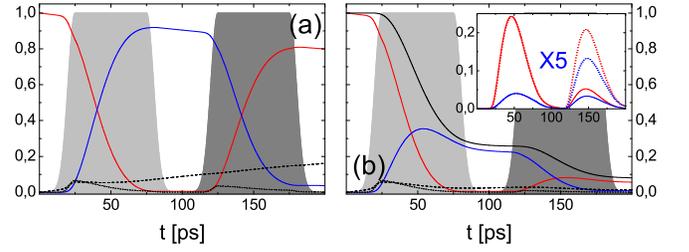}
\caption{(Color online). Simulated time evolution of the
unconditional (a) and conditional (b) density matrices under the
effect of two consecutive laser pulses, with frequencies $\omega_1$
(light gray area) and $\omega_2$ (dark gray). The curves show the
occupations of: state $ |1+\rangle $ (red line), state $ |2+\rangle
$ (blue), the trion manifold $|n\!>\!2\,\,\pm\rangle $ (black
dotted), and the ``$-$'' subspace (black dashed). The solid black
line in panel (b) corresponds to $\mathrm{Tr}\{\rho(t)\}$.  Figure
inset: photon occupations $ \langle a^{\dagger}_{\zeta} (t)
a_{\zeta} (t) \rangle $ for $ \zeta = + $ (red) and $ \zeta = - $
(blue, $\times 5$), in the conditional (solid lines) and
unconditional (dotted) cases.  The values of the parameters are: $
g_{1,2} = 0.2 , 0.15 \, $meV, $ \Omega_{1}^{(1)} (t_s) = 0.4 \,$meV,
$ \Omega_{2}^{(1)} (t_s) = 0.3 \, $meV, $ \kappa = 0.1 \,
$ps$^{-1}$, $ \Gamma_S = 1 \, $ns$^{-1}$, $ \Gamma = 2\, $ns$^{-1}$;
(b) $ \eta = 0.75 $.\vspace{-.8cm}} \label{fig2}
\end{center}
\end{figure}

To assess the performance of the proposed measurement strategy
[Fig.~\ref{fig1}(a)] one can take the signal output density matrix
for the spin \cite{ralph} as
$\rho_\mathrm{\,O}\!=\!\sum_\zeta\langle\hat{P}_\zeta \rangle
\mathrm{Tr}_{\mathrm{orb}} \{\sigma^{(\zeta)}_0\}/(2
\sum_\zeta\langle\hat{P}_\zeta \rangle )$, where the orthogonal
projectors $\hat{P}_\pm(T)=\hat{S}(T)[\hat{S}(T)\pm 1]/2 $
correspond to the events $N_+(T)\gtrless N_-(T)$. Here
$\hat{S}(T)\equiv \mathrm{sgn}[\hat{N}_+(T) - \hat{N}_-(T)]$ and $T$
is the total time over which the photocounts are integrated. The
signal input density matrix can be defined as
$\rho_\mathrm{\,I}\!\!=\!\!\mathrm{Tr}_{\mathrm{orb}}
\{\sum_{\zeta}\sigma^{(\zeta)}_0\rho_h \sigma^{(\zeta)}_0\}$. Then,
the quality of the measurement can be characterized by
$F^\mathrm{Min}_M(T)=\mathrm{Min}\{F[\rho_\mathrm{\,O}(T),
\rho_\mathrm{\,I}]\}|_{\rho_\mathrm{\,I}}$ \cite{ralph} where $F$ is
the square of the standard fidelity~\cite{nielsen}. The probability
of measurement ``failure'' ($N_+=N_-$) is given by $\langle
1-\hat{S}(T)^2 \rangle$. In all the regimes we will study below,
either the deviations from $H_\mathrm{eff}$ will be small enough to
guarantee that the emission of anticorrelated photons remains
improbable or the emission of more than one photon will have low
probability. It is straightforward to argue that under these
circumstances, it is permissible to redefine $\langle
\hat{P}_\pm\rangle \to \kappa\eta \int_0^T\! \mathrm{d}t\, \langle
a^{\dag}_{\pm} (t) a_{\pm}^{\vphantom{\dag}} (t) \rangle|_\eta$,
which is the probability that the first photon detected has
polarization $\sigma_\pm$.

We analyze now the physical limits for the measurement time $T$.
Naturally, there will be non-trivial spin dynamics neglected in
$H(t)$ -- e.g. spin decoherence -- that will set an upper limit for
$T$. On the other hand, a lower limit for $T$ is set by the
requirement that $F^\mathrm{Min}_M(T)$ and $\langle \hat{S}(T)^2
\rangle$ be close to unity. Within the above approximations we have
$F^\mathrm{Min}_M(T) = 1$ and the measurement can be considered
completed when the probability of not having detected a photon falls
below $\epsilon \ll 1$. From Eq.~(\ref{eq:mu}) it follows that this
probability evolves as $ \mathrm{Tr} \{ \mu (t) \} = e^{-\eta \kappa
P(t)}$. This implies $T=(-\ln\epsilon/\eta\tilde\Omega_0^2 y)^{1/2}
f^{-1}[y]$ with $y = -\kappa^2\ln\epsilon/\eta\tilde\Omega_0^2$.
Numerical optimization then yields $T_\mathrm{min}\approx
3(-\ln\epsilon/\eta)^{1/2}/\tilde\Omega_\mathrm{max}$ for
$\kappa_\mathrm{opt} \approx
\sqrt{\eta/\ln(1/\epsilon)}\tilde\Omega_\mathrm{max}$, with
$\tilde\Omega_\mathrm{max}$ a maximum value of $\tilde\Omega_0$
allowed by the conditions we discussed needed for the validity of
$H_\mathrm{eff}$. If $\kappa_\mathrm{opt}$ cannot be reached the
optimum is to take instead the lowest possible $\kappa$.
Experimental developments prompt us to consider as a typical
example: $\eta=0.75$, $\hbar\Delta=2$meV, $\hbar g_1=0.2$meV, $\hbar
\kappa= 0.05$meV and $g_1/g_2=\sqrt{3}$; which lead to $T\approx
200$ps for $\epsilon\sim 10^{-2}$, with
$\hbar\Omega_1^{(1)}(t_s)=0.4$meV.

The above analysis of the system's evolution and the resulting
timescales for $T$, that follow from Eq.~\ref{eq:mu}, rely on the
simultaneous switch-on (SSO) of the two lasers. An alternative
approach consists in applying an alternating sequence of
non-overlapping pulses with frequencies $\omega_1 , \omega_2 ,
\omega_1 , \omega_2 , \dots $ (see Fig.~\ref{fig2}). In this case,
each pulse triggers the emission of a single photon. We find that an
analysis based on $ H_\mathrm{eff} $, analogous to the one performed
for the SSO strategy, allows to establish $\kappa_\mathrm{opt}
\approx 2\tilde\Omega_\mathrm{max}$, and that $T_\mathrm{min} \sim
1/\eta\tilde\Omega_\mathrm{max}$ for the complete pulse sequence.
Thus, though the SSO strategy is preferable for low $\eta$, in the
range we will focus on ($\eta \gtrsim 0.7$) the timescales $T$ for
the two approaches are comparable. On the other hand, the ``pulsed''
scenario directly relates to photon-correlation experiments, that
allow to probe the spin's ``intrinsic'' dynamics.

In the following, we numerically solve the complete conditional
master equation for the QDM-MC system. The unitary part of the
evolution is induced by the Hamiltonian $H(t)$ (see Eq.~\ref{eq:H}).
The dissipative contribution, instead, is given by $
\mathcal{L}_C^{(+)}(\eta) +
  \mathcal{L}_C^{(-)}(\eta) +
  \mathcal{L}_T (t)         +
  \mathcal{L}_{S}             $,
where the collapse operators of the Liouvillian $\mathcal{L}_S$,
accounting for the spin-flip process, are
$\sqrt{\Gamma_S}\sigma_{\uparrow\downarrow}$ and
$\sqrt{\Gamma_S}\sigma_{\uparrow\downarrow}^\dag$, with
$\sigma_{\uparrow\downarrow}\equiv\sum_{n=1}^2 |n+\rangle\langle
n\!-\!|$. In Fig.~\ref{fig2} we plot the system's time evolution
under the effect of two consecutive laser pulses (shaded gray areas)
for both the unconditional [$\eta = 0$, panel (a)] and the
conditional case [$\eta \neq 0$, panel (b)]. The first pulse
essentially induces a population transfer from the initial state $
|1+\rangle $ (blue line) to $ |2+\rangle $ (red). The second Raman
transition, due to the following laser pulse, drives the hole back
to dot $L$. While the overall occupation of the excitonic manifold
(black dotted lines) is kept negligible throughout the process,
$\rho$ suffers a population leakage to subspace ``$-$'' (dashed
lines), which is responsible for the finite probability of emitting
a $\sigma_-$ photon (blue lines in the inset). We note that these
numerical simulations clearly support the approximations
underpinning the effective Hamiltonian $ H_\mathrm{eff} $. The
merits of the measure ultimately depend on the occupations of the
cavity mode. In particular, we find that the final probability of
having recorded no photocounts ($\mathrm{Tr}\{\rho\}|_{\eta\neq 0}$)
falls below $0.1$, while $\langle \hat{P}_+ \rangle= 0.71 \, (0.87)
$ and $\langle \hat{P}_- \rangle= 0.027 \, (0.048)$ after one pulse
(two pulses), yielding $ F^\mathrm{Min}_M = 0.963 \, (0.947) $. The
repetition of the measurement while decreasing
$\mathrm{Tr}\{\rho\}|_{\eta\neq 0}$, slightly worsens the fidelity.
This is because the repetition time is not sufficiently short
compared to $ 1 / \Gamma_S $.

\begin{figure}[tbp]
\begin{center}
\includegraphics[width=1.\columnwidth]{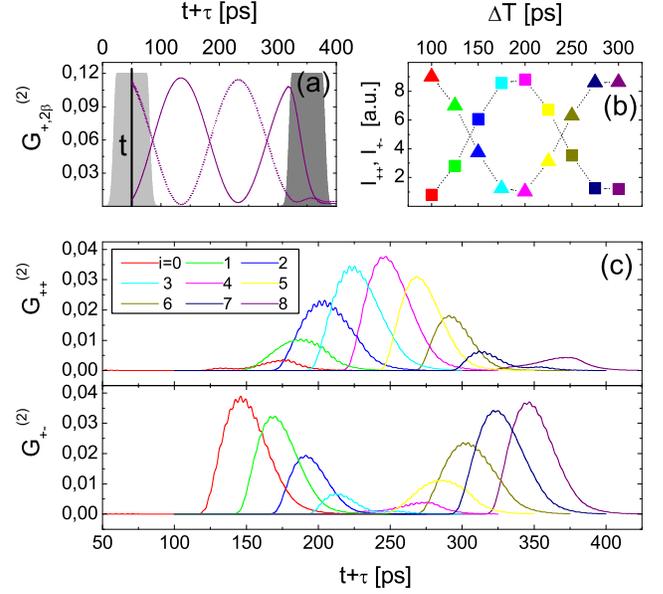}
\caption{(Color online). (a) Correlation functions $
G^{(2)}_{+,2\pm} ( t , t + \tau ) $.  The pulse from laser 1 (light
gray) and the detection of a $\sigma_+$ photon at time $t$ (vertical
black line) initialize the spin state, setting in damped
oscillations between states $ | 2 + \rangle $ (dotted curve) and $ |
2 - \rangle $ (solid).  The pulse from laser 2 (dark gray, $ \Delta
T = 300\, $ps), probes the spin evolution.  (b) Time integrals of $
G^{(2)}_{++} $ (squares) and $ G^{(2)}_{+-} $ (triangles), for
different values of the delay between the two pulses: $ \Delta T =
(100 + 25\, i)\, $ps, with $ i = 0,\dots ,8$ (see the legend). (c)
Correlation functions $ G^{(2)}_{+\beta} ( t , t + \tau ) $, for $
\beta = + $ (upper panel) and $ \beta = - $ (lower panel), as a
function of $ t + \tau $ ($t = 50$~ps). $B_x =0.5$~T.
\vspace{-.8cm}} \label{fig3}
\end{center}
\end{figure}

The non-destructive nature of our measurement scheme turns it into
an ideal means to probe the spin's dynamics.  In particular, the
polarization correlations between two photons detected at times $ t
$ and $ t + \tau $ can be used to investigate the evolution that the
spin undergoes in such time interval.  As above, the system's state
is driven by a sequence of two non-overlapping laser pulses, with
frequencies $\omega_1$ and $\omega_2$ [Fig.~\ref{fig3}(a)].  The
first pulse (light gray area) induces a Raman transition which
displaces the hole from dot $L$ to dot $S$. The cases where a
$\sigma_+$ photon is detected at a given time $t$ (vertical black
line) are post-selected: this first measurement (approximately)
projects $ \rho $ onto the $|2+\rangle$ state, thus initializing the
spin to a pure state.  A tunable time-interval $ \Delta T $ follows,
during which the spin freely evolves under the effect of the
spin-flip process, and eventually of an applied magnetic field
$\mathbf{B}$.  The corresponding time-evolution, conditioned upon
having detected a photon at time $t$, is given by the second-order
correlation functions $G^{(2)}_{ \zeta , n\beta } ( t , t + \tau ) =
\langle \, a_{\zeta}^{\dagger} ( t ) \, \sigma_{nn}^{( \beta )} ( t
+ \tau ) \, a_{\zeta} ( t ) \, \rangle $, with $ \zeta , \beta = \pm
$. Finally, the second laser pulse (dark gray) probes the spin
state, while displacing the carrier back to dot $L$.  If the
magnetic field is applied in the $z$ direction, the polarization
correlations between the first and the second detected photons,
given by $ G^{(2)}_{ \zeta \beta } ( t , t + \tau ) = \langle \,
a_{\zeta}^{\dagger} ( t ) \, a_{\beta}^{\dagger} ( t + \tau ) \,
a_{\beta } ( t + \tau ) \, a_{\zeta} ( t ) \, \rangle $, only
reflect the effect of $ \mathcal{L}_S $, allowing to infer the value
of $ T_1 = 1 / \Gamma_S $.  If $\mathbf{B} \!\!\parallel \!\!
\hat{x} $ instead~\cite{Zeeman}, $J_z$ is no longer a constant of
motion of $H(t)$, and its time-evolution will consist of damped
oscillations between the states $|2+\rangle$ and $|2-\rangle$ (solid
and dotted curves, respectively).  Due to the high fidelity of the
initializing measurement, the initial conditions do not play here a
crucial role. As the energy splitting induced by $\mathbf{B}$ is
small compared to $k_B T$, we take $ \rho_h = ( | 1+ \rangle\langle
1\!+\! | + | 1- \rangle\langle 1\!-\! | ) / 2 $. In
Fig.~\ref{fig3}(c) we plot $ G^{(2)}_{ \zeta \beta } ( t , t + \tau
) $, for $\zeta = + $ and $\beta = \pm$ (upper and lower panels,
respectively), and for different values of $ \Delta T $. The time
integrals of these functions $ I_{\zeta\beta} = \int G^{(2)}_{\zeta
\beta } ( t , t + \tau ) \, d\tau $ clearly show an oscillatory
behavior as a function of $ \Delta T $ [Fig.~\ref{fig3}(b)]. The
free damped oscillations we observe reflect the decay of the initial
coherence between the eigenstates of $J_x$, and thus would allow to
infer the $T_2$ for a transverse field.

In conclusion, we have proposed an all-optical robust scheme to
perform a QND measurement of a single hole spin in sub-nanosecond
timescales. Furthermore, we have pointed out how in the presence of
a static magnetic field photon correlation experiments would allow
to study the spin decoherence. Beyond measurement, the entanglement
between the carrier and the photon could enable generation of EPR
pairs. In the case of correlations with the phase quadratures
[Eq.(\ref{eq:sstate})] one could also envisage the generation of
Schr\"{o}dinger cat states of the emitted light. Finally, the same
system could be operated in a continuous measurement regime with the
spin-flip processes inducing quantum jumps in the output.

We thank J.C. Cuevas and J. Eschner for discussions. Work supported
by the Spanish MEC under the contracts MAT2005-01388 and
NAN2004-09109-C04-4, by CAM under Contract S-0505/ESP-0200, and by
the EU within the RTN's COLLECT and CLERMONT2.

\end{document}